\documentclass[superscriptaddress,twocolumn,showpacs,preprintnumbers,amsmath,amssymb]{revtex4}

\usepackage{epsfig}
\usepackage{dcolumn}

\usepackage[latin1]{inputenc}
\usepackage{amsmath}
\usepackage{psfrag}
\usepackage{color}
\newcommand{\chem}[1]{\ensuremath{\mathrm{#1}}}
\newcommand{\un}[1]{\ensuremath{\unskip\,\mathrm{#1}}}

\begin{document}


\title{Noise-assisted spike propagation in myelinated neurons}

\author{Anna Ochab-Marcinek }
\affiliation{Institut f\"ur Physik, Universit\"at Augsburg, Universit�tsstr. 1, 86159 Augsburg, Germany}
\affiliation{M. Smoluchowski
Institute of Physics, Jagellonian University, ul. Reymonta 4, 30-059
Krak\'ow, Poland}

\author{Gerhard Schmid}
\affiliation{Institut f\"ur Physik, Universit\"at Augsburg, Universit�tsstr. 1, 86159 Augsburg, Germany}

\author{Igor Goychuk}
\affiliation{Institut f\"ur Physik, Universit\"at Augsburg, Universit�tsstr. 1, 86159 Augsburg, Germany}

\author{Peter H\"anggi}
\affiliation{Institut f\"ur Physik, Universit\"at Augsburg, Universit�tsstr. 1, 86159 Augsburg, Germany}
\date{\today}

\begin{abstract}
We consider noise-assisted spike propagation in myelinated axons within
a multi-compartment stochastic Hodgkin-Huxley model. The noise originates from
 a finite number of ion channels in each node of Ranvier. For the subthreshold internodal
electric coupling, we show that (i) intrinsic noise removes the
sharply defined threshold for spike propagation from node to node, and
(ii) there exists an optimum number of ion channels which allows for
the most efficient signal propagation and it corresponds to the actual physiological values.

\end{abstract}



\maketitle
\section{\label{sec:intro}Introduction}

In many vertebrates, the propagation of nerve impulses is mediated
by myelinated axons. These nerve fibers are composed of active
zones, the so-called nodes of Ranvier, where ion channels are
accumulated, and passive fragments which are electrically isolated
(myelinated) from the surrounding electrolyte solution. The myelin
sheath presents a crucial evolutionary innovation. In a process
called saltatory conduction, the neural impulse propagates from node
to node more rapidly than it propagates in  unmyelinated axons of
equal diameter~\cite{Koch,Keener}. Myelination not only helped
Nature to increase the signal propagation speed but also to
drastically reduce the metabolic cost of neural computation. Indeed,
with a single neural impulse (of order of 1 ms), each sodium channel
transfers up to  $10^4$ sodium ions into the cell which then should be pumped back to
restore the appropriate steady-state electrochemical potential. To
transfer three sodium ions, the corresponding \chem{Na}-\chem{K}
pump  hydrolyzes  one ATP molecule ~\cite{Keener}, which  thus
yields the estimate of about $3\cdot 10^3$ ATP molecules per ion
channel per neural spike.  A too large number of channels imposes an
extremely high metabolic load (the brain of the reader is just now
consuming about 10\% of the body's metabolic budget, which, per one
kilogram of mass, is more than the muscles use when active). More
elaborate estimates confirm that even a small cortical cell needs,
in a long run, at least $10^7$ ATP molecules per one
neural spike~\cite{Laughlin,Bezrukov}.\\

In myelinated neurons, yet another problem emerges: there is a
threshold present for the electrical coupling between the nodes of
Ranvier. Deterministic cable equation models predict that if the
internodal distance is too large, the coupling becomes too small and
the signal propagation consequently  fails ~\cite{Keener}. However,
due to a finite number of channels (of the order of $10^4$),
intrinsic noise is inevitably present in the nodes of
Ranvier~\cite{Lecar}. The deterministic models can only mimic the
behavior of a very large number of ion channels possessing a
negligible intrinsic noise intensity; in contrast,   real neurons
 tend to minimize the number of ion channels because of  energetic costs. On
the other hand, too few  channels may trigger random, parasitic
spikes or induce spike suppression by strong fluctuations, making
the signal transmission too noisy and thus unreliable. The questions
we investigate in this paper within a simplified stochastic model of
signal transmission in a myelinated neuron, are as follows: can
channel noise soften the propagation threshold and facilitate the
signal propagation which would not occur in the deterministic case?
Does an optimum size of the channel cluster exist, which in turn
yields  a most efficient, noise-assisted
propagation?\\

The statement of this objective shares 
features similar in spirit to noise supported wave propagation in
subexcitable media (see Ref.~\cite{sagues} and references therein)
and noise in excitable spatiotemporal systems
(see Ref.~\cite{lindner}). The influence of intrinsic noise on
membrane dynamics was
studied in the context of intrinsic stochastic resonance, see in Refs.~\cite{Schmid,Jung}, and 
synchronization of ion channel clusters \cite{Zeng}. 
A new aspect emerging from the present study is a possible
interpretation of the neural spike transmission as a delayed
synchronization phenomenon occurring in the chain of active
elements~\cite{Hennig1,Hennig2,Glatt}. 


\begin{figure*}[t]
  \centering
  \epsfig{figure=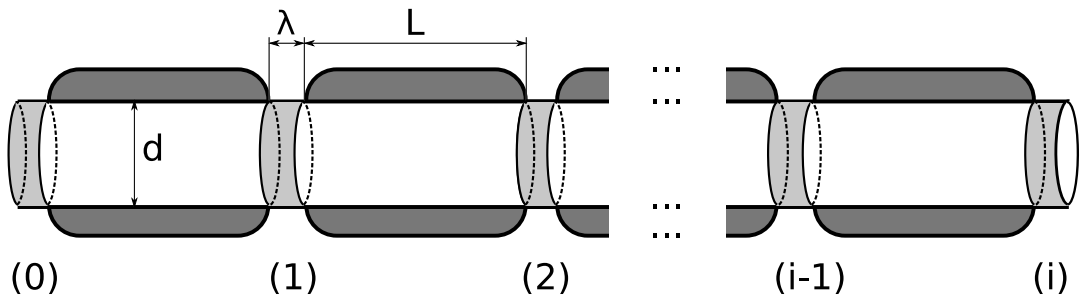,scale=0.7}\\
  \caption{Sketch of a myelinated axon: The axonal cell membrane in
    the nodes of Ranvier (depicted in light-grey) contains a high ion
    channel concentration. The remaining segments are wrapped in the
    myelin sheath (dark-grey). Spike propagation along the axon occurs
    in a saltatory manner.}  
  \label{fig:comp_mod}
\end{figure*}

\section{\label{sec:hh}The model}
\subsection{\label{subsec:comp_mod} Compartment modelling for myelinated
axons}

As an archetypical  model for signal transmission along a myelinated
axon of a neuronal cell we consider a compartmental stochastic
Hodgkin-Huxley model (see Sec.~\ref{subsec:hh},
\ref{subsec:hh_stoch}). In contrast to unmyelinated axons, where ion
channels are homogeneously distributed within the membrane, the
myelinated axon consists of alternating sections where the ion
channels are densely accumulated (nodes of Ranvier), and regions
with very low ion channel density, encased in multiple layers of a
highly resistive lipid sheath called myelin. The nodes of Ranvier
play the role of {\it active} compartments in our model, while the
electrically neutral myelinated segments constitute {\it passive}
compartments (see Fig.~\ref{fig:comp_mod}). The typical internodal
distance $L$ is about 1-2 mm while the length $\lambda$ of the nodes
of Ranvier is in the micrometer range \cite{Koch}. 

The generalization of the original Hodgkin-Huxley model is related to
the influence of the channel noise which results from the randomness of the ion channel gating.
The membrane potential at each particular node of Ranvier is assumed to be constant across the
whole node area and is characterized by $V_{i}$, $i=0,1,2,....N-1$ where $N$
is the total number of nodes~\cite{Koch,Keener}.

Supposing  total electrical neutrality of the {\it passive} regions,
the electrical properties of a myelinated axon are modeled by a
linear chain of diffusively coupled active elements. The dynamics of
the membrane potential is  given by:
\begin{multline}
  \label{eq:comp_mod}
  C \frac{\mathrm d}{\mathrm{d} t} V_{i} = I_{i,
    \mathrm{ionic}}(V_{i},t) + I_{i, \mathrm{ext}}(t) \\+ I_{i, \mathrm{inter}}\, , \quad \text{for
  } i=0,1,2,....N-1\, ,
\end{multline}
where $C$ denotes the axonal membrane capacity per unit area. $I_{i, \mathrm{ionic}}(V_{i},t)$ is the
ionic membrane current (per unit area) within the $i$th node of Ranvier. $I_{i,
  \mathrm{ionic}}(V_{i},t)$ depends on time $t$ and on the membrane potential $V_i$ in the specific
node ${i}$ only, and is described by a neuronal membrane model which
we subsequently treat within a coupled  stochastic Hodgkin-Huxley
setup). $I_{i,\mathrm{ext}}(t)$ describes an external current
per unit area applied on the $i$th node. In our setup we apply the
stimulus, i.e. a constant current, on the first node only, bringing
it into a dynamical regime of periodic firing~\cite{izhikevich2000}.
If the coupling between nodes is sufficiently strong, the action
potentials may start propagating along the axon. The coupling to the
next-neighboring nodes of Ranvier is achieved by inter-nodal
currents (per unit area) $I_{i, \mathrm{inter}}$, given as:
  \begin{align}
  \label{eq:internodal-currents}
    I_{i, \mathrm{inter}} =
    \begin{cases}
  \kappa \left( V_{i+1} - V_{i}\right) & \text{for }i=0\, ,\\
  \kappa \left( V_{i-1} - V_{i}\right) & \text{for }i=N-1\, ,\\
\kappa
  \left( V_{i-1}-2 V_{i} + V_{i+1}\right) & \text{elsewhere.}
    \end{cases}
  \end{align}
The coupling parameter $\kappa$ 
depends, among other things, on the length $L$
of the internodal {\it
  passive} segment of the axon as well as on the resistivity 
of the extracellular and
intracellular medium, and serves as a control parameter in our studies
\cite{Koch}. It depends also on the ratio of the node's diameter $d$
and its length $\lambda$. Typically, the node's diameter is two orders of magnitude
smaller than its length \cite{Keener}.

Note that the Eq.~\eqref{eq:comp_mod} just presents the Kirchhoff
law for an electrical circuit made of membrane capacitors, variable
nonlinear membrane conductances and internodal conductances
$\kappa$.

\subsection{\label{subsec:hh} Hodgkin-Huxley modeling}

According to the standard Hodgkin-Huxley model~\cite{Hodgkin1952} the
ionic membrane current reads:
\begin{multline}
  \label{eq:voltage-equation}
  I_{i}(V_{i}) = -G_{\chem{K}}(n_{i})\ (V_{i}-E_{\chem{K}})\\
 -G_{\chem{Na}}(m_{i},h_{i})\ ( V_{i} -  E_{\chem{Na}})
  -G_{\chem{L}}(V_{i} -  E_{\chem{L}})\, ,
\end{multline}
with the potassium and sodium conductances per unit area given by:
\begin{equation}
  \label{eq:conductances-hodgkinhuxley}
  G_{\chem{K}}(n_{i})=g_{\chem{K}}^{\mathrm{max}}\  n_{i}^{4} , \quad
  G_{\chem{Na}}(m_{i},h_{i})=g_{\chem{Na}}^{\mathrm{max}}\ m_{i}^{3} h_{i}\, .
\end{equation}
In Eq.~(\ref{eq:voltage-equation}), $V_{i}$ denotes the membrane
potential at the $i$th node of Ranvier. Furthermore, $E_{\chem{Na}}$, $E_{\chem{K}}$ and
$E_{\mathrm{L}}$ are the reversal potentials for the potassium,
sodium and leakage currents, correspondingly. The leakage
conductance per unit area $G_{\mathrm{L}}$ is assumed to be constant. The
parameters  $g_{\chem K}^{\mathrm{max}}$ and
$g_{\chem{Na}}^{\mathrm{max}}$ denote the maximum potassium and
sodium conductances per unit area, when all ion channels within the corresponding
node are open. The values of these parameters are collected in the
Table~\ref{tab:params} \cite{note}. Note that in the nodal membrane
 the conductances of {\it open}
ion channels are supposed to be Ohmic-like, i.e. the nonlinearity
derives from their gating behavior only. Moreover, formulating the
Eqs.~\eqref{eq:voltage-equation} and
\eqref{eq:conductances-hodgkinhuxley}, we implicitly assumed for
simplicity that the different axonal nodes are kinetically
identical, i.e. the number of sodium and potassium ion channels is
constant for each node. As a consequence, the maximum potassium and
sodium conductances are identical constants for every node of
Ranvier.

The gating variables $n_{i},\ m_{i}$ and $h_{i}$, cf.
Eqs.~\eqref{eq:voltage-equation} and
\eqref{eq:conductances-hodgkinhuxley}, describe the probabilities of
opening the gates of the specific ion channels in the $i$th node upon
the action of activation and inactivation particles. $h$ is the
probability that the 1 inactivation particle has not caused the
$\chem{Na}$ gate to close. 
$m$ is the probability that 1 of the 3 required activation particles
has contributed to the activation of the $\chem{Na}$ gate. Similarly,
$n$ is the probability of the $\chem{K}$ gate activation by 1 of the 4
required activation particles. Assuming gate independence, the factors
$n_{i}^{4}$ and $m_{i}^{3}\ 
h_i$ are the mean portions of the open ion channels within a membrane
patch. The dynamics of the gating variables are determined by
voltage-dependent opening and closing rates $\alpha_{x}(V)$ and
$\beta_{x}(V)\; (x=m,h,n)$ taken at  $T=6.3 \un{^{\circ} C}$. They
depend on the local membrane potential $V$ and read (with numbers
given in units of $[\text{m} V]$)~\cite{Hodgkin1952,goychukpnas}:
\begin{subequations}
\label{eq:gating}
\begin{align}
  \label{eq:rates-m}
  \alpha_{m}(V) &= 0.1 \, \frac{V+40}{1 - \exp\left\{-\, (V+40)/10
    \right\}}\, , \\
  \beta_{m}(V)  &= 4\, \exp\left\{ -\, (V+65)/18 \right\}\, , \\
  \alpha_{h}(V) &= 0.07  \, \exp\left\{ -\, (V+65)/20 \right\}\, , \\
  \beta_{h}(V)  &= \frac{1}{1 + \exp\left\{-\, (V+35)/10 \right\} }\, ,  \\
  \alpha_{n}(V) &= 0.01\, \frac{V + 55}{1 - \exp \left\{-\, (V + 55)
      / 10 \right\}}\, ,\\
  \label{eq:rates-n}
  \beta_{n}(V)  &= 0.125 \, \exp\left\{ -\, (V+65)/80 \right\}\, .
\end{align}
\end{subequations}
The dynamics of the mean fractions of open gates reduces in the standard Hodgkin-Huxley
model to relaxation dynamics:
\begin{equation}
  \label{eq:gates}
  \frac{\mathrm{d}}{\mathrm{dt}} x_{i} =
  \alpha_{x}(V_{i})\ (1-x_{i})-\beta_{x}(V_{i})\ x_{i}, \quad x=m,h,n\, .
\end{equation}
Such an approximation is valid for very large numbers of
ion channels and whenever fluctuations around their mean values are negligible.

%

\subsection{Stochastic generalization of the Hodgkin-Huxley model \label{subsec:hh_stoch}}

Because the size of the nodal membrane is finite there necessarily occur
fluctuations of the number of open ion channels. This is
due to the fact that ion channel gating between open and closed
state is random. Accordingly, for finite size membrane patches like
the considered nodes of Ranvier, there are fluctuations of the
membrane conductance which  give rise to {\it spontaneous} action
potentials~\cite{Koch,Schmid,Jung,Chow1996,Tuckwell1987,Fox1994} and the references therein.

The number of open gates undergoes a birth-and-death-like process.
The corresponding master equation can readily be written down. The
use of a Kramers-Moyal expansion in that equation results in a
corresponding Fokker-Planck equation which provides a diffusion
approximation to the discrete dynamics~\cite{Tuckwell1987,Fox1994}.
The corresponding multiplicative noise Langevin equations
\cite{HanggiThomas_a} then read:
\begin{subequations}
\begin{align}
  \label{eq:stochasticgates}
  \frac{\mathrm{d}}{\mathrm{dt}} x_i =
  \alpha_{x}(V_{i})\ (1-x_i)-\beta_{x}(V_{i})\ x_i + \xi_{i, x}(t)\, ,
\end{align}
with $x=m,h,n$. Here, $\xi_{i, x}(t)$ are independent Gaussian white
noise sources of vanishing mean and vanishing cross-correlations.
For an excitable node with the nodal membrane size ${\cal A}$ the
non-vanishing noise correlations take the following form:
\begin{widetext}
\label{eq:noisecor}
\begin{align}
  \label{eq:correlator-a}
  \langle \xi_{i, m}(t)\, \xi_{i, m}(t') \rangle &=
  \frac{1}{{\cal A} \rho_{\chem{Na}}}\
   [\alpha_{m}(V_{i})\, (1-m_i) + \beta_{m}(V_{i})\, m_i]\ \delta(t -t')\, ,  \\
  \label{eq:correlator-b}
  \langle \xi_{i, h}(t)\, \xi_{i, h}(t') \rangle &=  \frac{1}{{\cal A} \rho_{\chem{Na}}}\
  [ \alpha_{h}(V_{i})\, (1-h_i) + \beta_{h}(V_{i})\, h_i]\ \delta(t -t')\, , \\
  \label{eq:correlator-c}
  \langle \xi_{i, n}(t)\, \xi_{i, n}(t') \rangle &=  \frac{1}{{\cal A} \rho_{\chem{K}}}\
  [ \alpha_{n}(V_{i})\, (1-n_i)+
    \beta_{n}(V_{i})\, n_i ]\ \delta(t -t')  \, ,
\end{align}
\end{widetext}
\end{subequations}
with the ion channel densities being $\rho_{\chem{Na}}$ and
$\rho_{\chem{K}}$ (see Table~\ref{tab:params}). The stochastic Eq.
(\ref{eq:stochasticgates}) replaces Eq.
(\ref{eq:gates})~\cite{Schmid2006}. Remarkably, the nodal
membrane size ${\cal A}$, which is the same for all nodes, solely
influences the intrinsic noise strength.  Note that the correlations
of the stochastic 
forces in these Langevin  equations contain the corresponding
state-dependent variables and, being an approximation to the full
master equation dynamics, thus should be interpreted in the $\rm
It\hat o$ sense~\cite{HanggiThomas}.

\onecolumngrid
\begin{center}
\begin{table}[t]
{\footnotesize
\begin{tabular}{lllll}
\hline
\hline
\textbf{Stochastic Hodgkin-Huxley model}&&&&\\
\hline
Membrane capacitance per unit area \ &$C$& = &1 &\un{\mu F/cm^2} \\
\hline
Reversal potential for Na current \ & $E_\chem{Na}$& = &$50$ &\un{mV} \\
Reversal potential for K current \ & $E_\chem{K}$& = &$-77$ &\un{mV} \\
Reversal potential for leakage current \ & $E_\chem{L}$& = &$-54.4$ &\un{mV} \\
\hline
Leakage conductance per unit area\ & $G_L$&=&0.25 &$\un{mS/cm^2}$ \\
Maximum Na conductance per unit area\ & $g^{\mathrm{max}}_{\chem{Na}}$&=&120 &$\un{mS/cm^2}$ \\
Maximum K conductance per unit area\ & $g^{\mathrm{max}}_{\chem{K}}$&=&36&$\un{mS/cm^2}$ \\
\hline
Node area \ & ${\cal A}$& varying & & $[\un{\mu m^2}]$\\
Na channel density \ & $\rho_{\chem{Na}}$&=& 60 &$\un{\mu m^{-2}}$ \\
K channel density \ & $\rho_{\chem{K}}$&=&18&$\un{\mu m^{-2}}$ \\
Number of Na channels \ & $N_{\chem{Na}}$ & = & $\rho_{\chem{Na}} \
{\cal A}$&\\
Number of K channels \ & $N_{\chem{K}}$ & = & $\rho_{\chem{K}} \ {\cal
A}$&\\
\hline
\hline
\textbf{Axonal chain model}&&&&\\
\hline
Inter-nodal conductance per unit area\ &$\kappa$& = &  0.065  &$\un{mS/cm^2}$ \\
Number of nodes \ & $N$ & = & 10 &\\
\hline
\hline
\textbf{Simulation parameters}&&&&\\
\hline
External current per unit area, node 0 \ & $I_{0, \mathrm{ext}}$& = & 12 &$\un{\mu A/cm^2}$\\
External current per unit area, other nodes \ & $I_{i=1..N-1, \mathrm{ext}}$ & = & 0 &$\un{\mu A/cm^2}$ \\
\hline
Simulation time & $T$ & = &$ 3 \cdot 10^5$ &\un{ms} \\
Simulation time step & $\mathrm{d}t $ &= & $0.002$ &\un{ms}\\
\hline
\hline
\textbf{Initial values for each node} ($i=0,..,N$)&&&& \\
\hline
Voltage \ & $V_{i}$ & =& -59.9 & mV \\
Inactivation probability for the \chem{Na} gate
&$h_{i}$&=& 0.414 \\
Activation probability for the \chem{Na} gate
&$m_{i}$& = &0.095 \\
Activation probability for the \chem{K} gate.
&$n_{i}$ & = & 0.398 \\
\hline
\end{tabular}
}
\caption{\label{tab:params} Model and simulation parameters}
\end{table}
\end{center}
\twocolumngrid

\section{\label{subsec:opt}Optimization of the signal propagation}

We next study the model of a myelinated axon in which one node of
Ranvier is continuously excited to spontaneous spiking by an
external supra-threshold current, whereas other nodes are initially
in the resting state and the inter-nodal conductance $\kappa$ is too
low for the spikes to propagate to the neighboring nodes in the deterministic model (notice the
arrow in Fig.~\ref{fig:kappa}). The fluctuations of the potential
caused by the randomly opening and closing channels can be large
enough to help the spike to overcome the large  inter-nodal
resistance and to make the saltatory conduction between nodes
possible.

\subsubsection*{Deterministic limit}

We start out with the deterministic situation using a finite number of
Ranvier nodes, i.e. the channel noise level is set to zero, which is
formally achieved in the limit ${\cal A} \to \infty$. In order to provide
some insightful explanation on the dependence of saltatory spike
propagation on the coupling parameter, we numerically integrated the
dynamical system, see. Eqs.~\eqref{eq:comp_mod}-\eqref{eq:gates}
with the parameters given in Tab.~\ref{tab:params}. In order to
achieve equilibration along the whole chain we do not apply the
constant current on the first node $I_{0,\mathrm{ext}}$ from the
beginning. Instead, we use the following protocol: (i) we initially
allow for equilibration of every individual node by integrating over
$100\un{ms}$ with $\kappa = 0$; (ii) in a second step we switch on
the coupling; (iii) finally after integrating over $150\un{ms}$,
i.e. under stationary conditions, we apply a constant current
stimulus of $I_{0,\mathrm{ext}}=12 \un{\mu A/cm^{2}}$ on the
first node only. Due to the supra-threshold stimulus~\cite{Schmid}
on the first node, the dynamics of the membrane potential of the
initial node $V_{0}$ exhibits a limit cycle, resulting in the
periodic generation of action potentials. The numerically
obtained spike trains $V_{i}$ for $i=0,..,9$ allow for studying the
propagation of action potentials along the linearly coupled
chain. Using a threshold value of $V_\mathrm{th}=20 \un{mV}$ enables the
detection of an action potential (also called spike or firing event) in the particular spike train of the individual
nodes. Since we are interested in the successful transmission
along the whole chain, the spiking at the terminal node is analyzed
and the number of spikes at the terminal node is related to the number
of initiated spikes at the initial node. This defines the transmission
reliability. 

\begin{figure}[t]
\begin{center}
\epsfig{figure=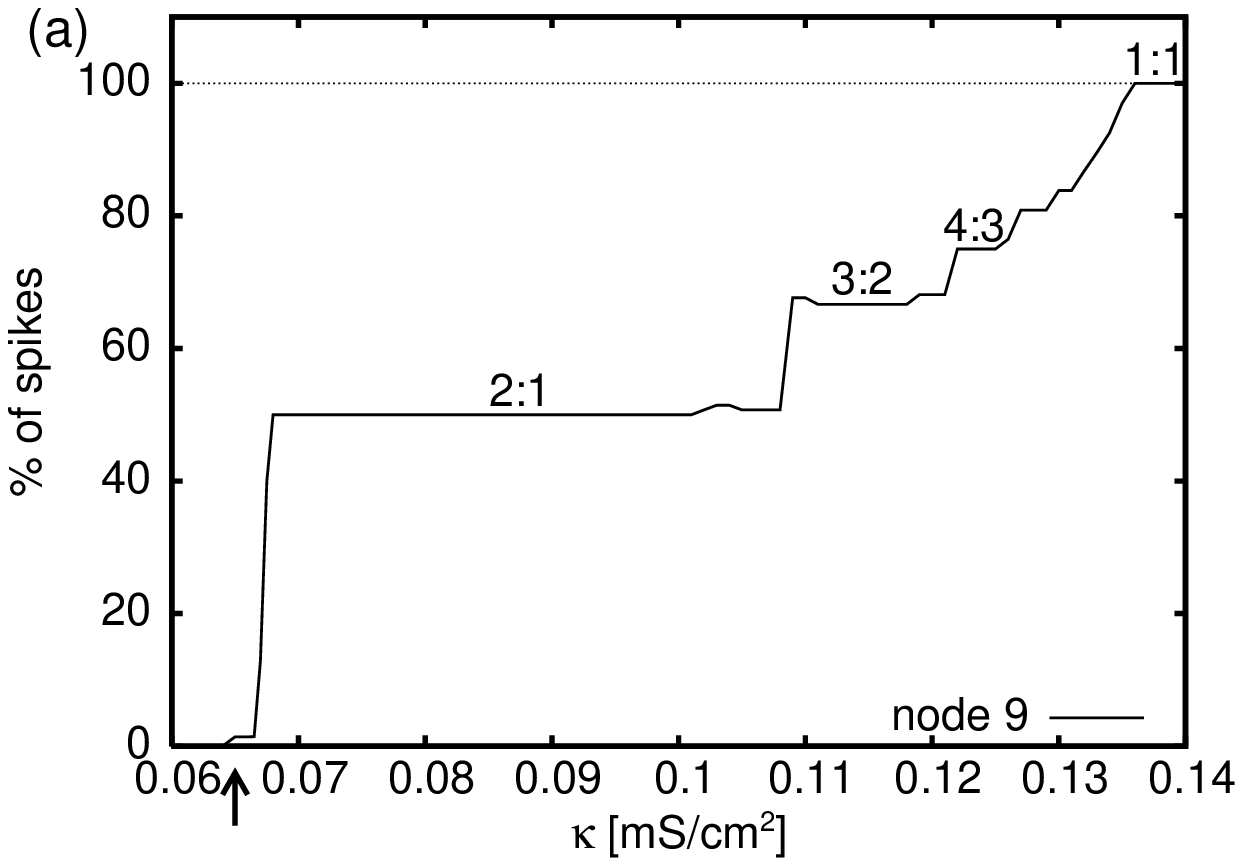,scale=0.6}\\
\vspace{0.5cm} \epsfig{figure=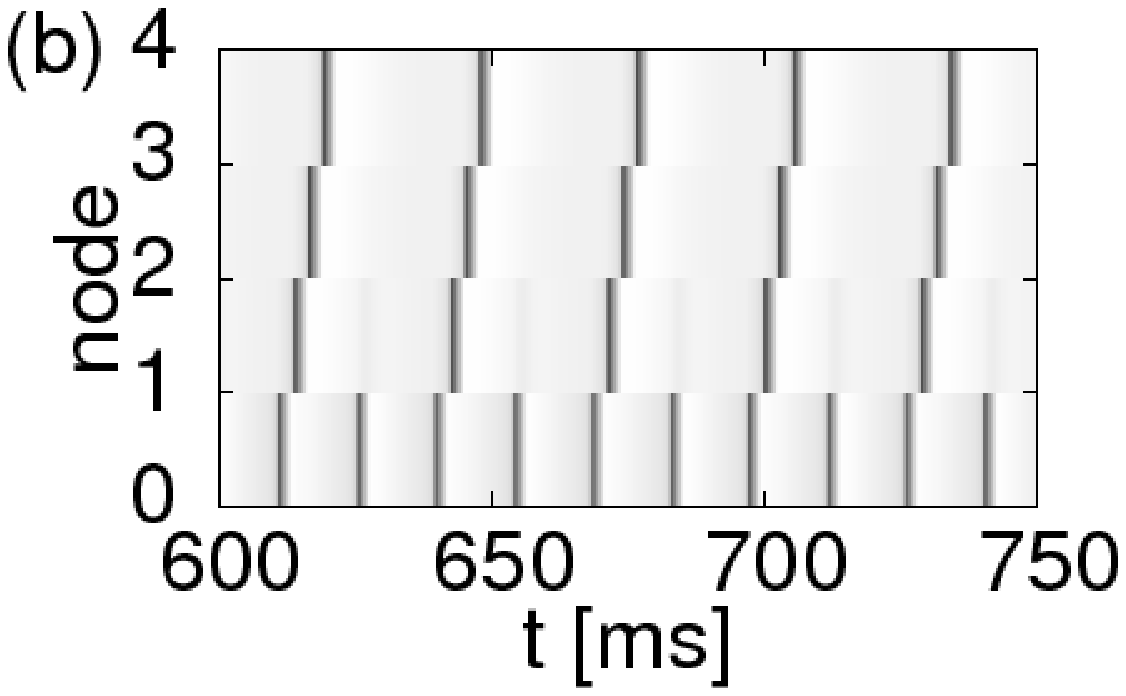,scale=0.35}
\epsfig{figure=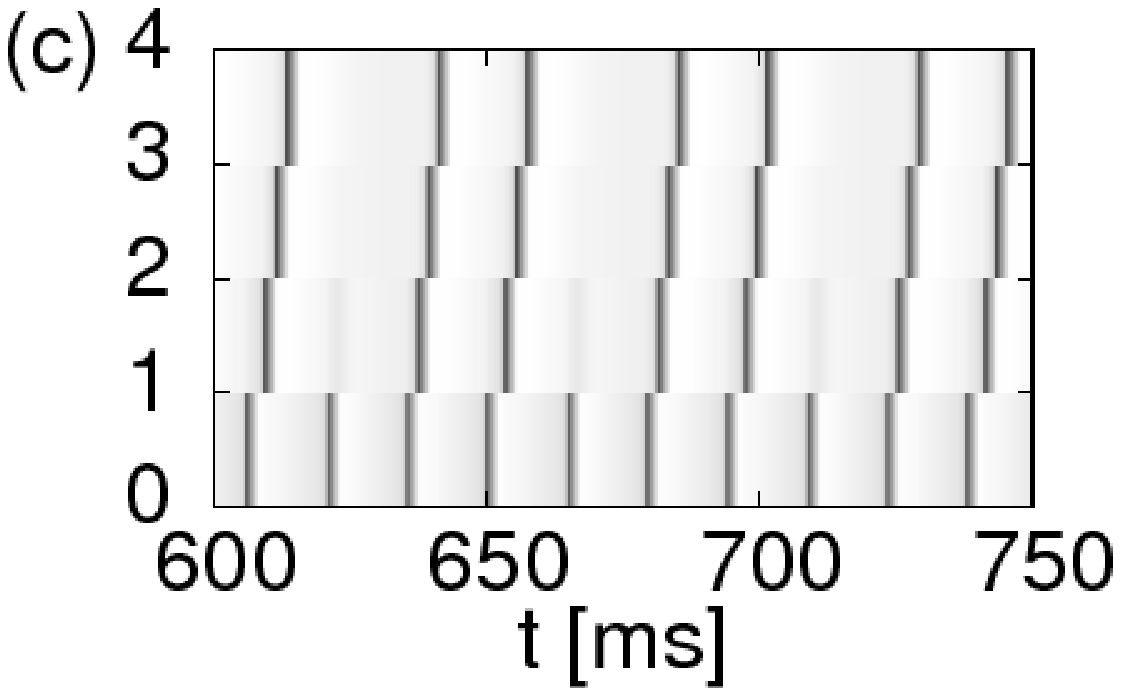,scale=0.35}
\epsfig{figure=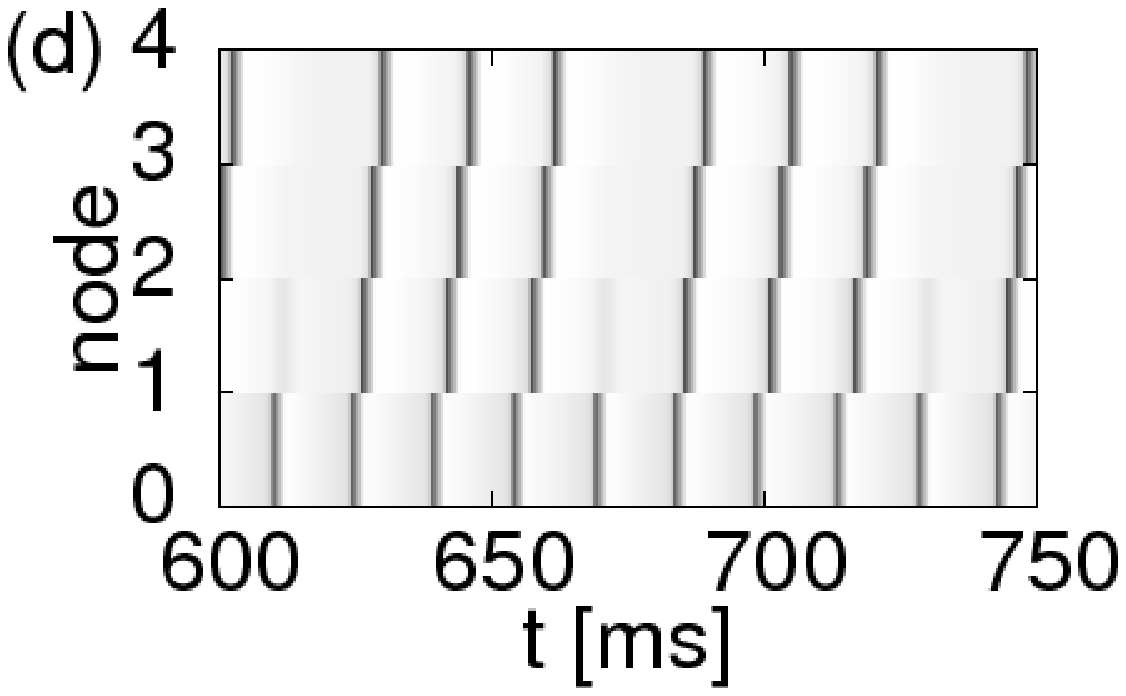,scale=0.35}
\epsfig{figure=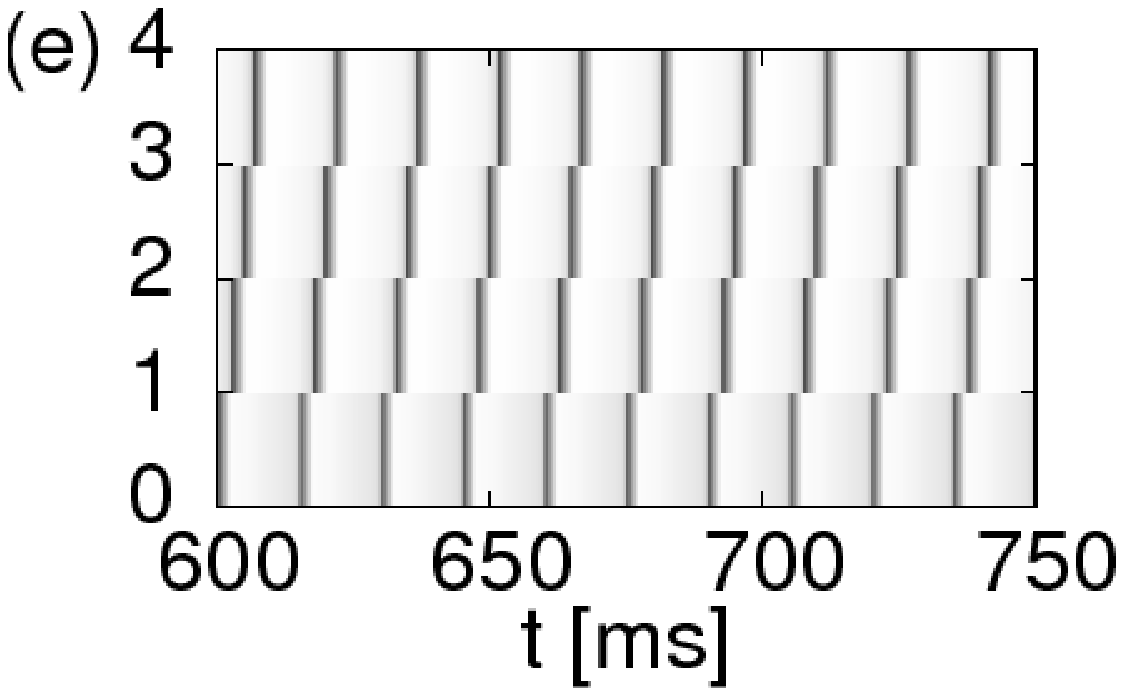,scale=0.35} \caption{\label{fig:kappa}
Deterministic spike propagation in a myelinated axon containing
$N=10$ Ranvier nodes (no channel
  noise): (a) The ratio of spikes generated in the node 0
  during an interval of 1000 \un{ms} which propagate up to the node 9
  (solid line), as a function of  the value of the inter-nodal coupling strength
  (i.e. the conductance)
  $\kappa$. The arrow indicates the subthreshold value of
  $\kappa=0.065 \ \un{mS/cm^2}$ chosen for further simulations of
  the influence of channel noise on spike propagation. The stepwise
  shape of the graph depicts the different transmission patterns: (b)
  2:1 ; (c) 3:2 ; (d) 4:3 and (e) 1:1 .
For example, for the ratio of 2:1, every second spike which is
generated at $n=0$ propagates. (Small irregularities in step levels
occur as a result
  of a finite counting statistics of propagating spikes.)}
\end{center}
\end{figure}

For the chosen coupling
strength $\kappa$, the excitation propagates along the chain of
given $N=10$ nodes of Ranvier. 
This particular choice of the number of
nodes is somewhat arbitrary, although physiologically realistic. Note
also, that these obtained results are robust and are typical for larger
node numbers as well (not shown).
Obviously, for too small coupling
parameters no excitation proceeds to the neighboring node ``1'' and,
at the end, will not reach the terminal node ``9''. In this case,
 we observe that spike propagation fails. In the opposite limit,
i.e. for large coupling constants $\kappa$, each  action potential
travels along the chain and arrives at the terminal node ``9''. Then
the spike transmission efficiency of $100\%$ is achieved. In this situation, the
dynamics of the terminal node is  synchronized  with a delay with
the initial node, i.e. $V_{9}(t) = V_{0}(t-\tau)$, where the delay
time $\tau$ accounts for the finite propagation speed. Most
interestingly, however, there occurs no sharp transition between
those two regimes of $0\%$ and $100\%$ signal transmission (see
Fig.~\ref{fig:kappa}). For $\kappa_{c1}=0.0665 \un{mS/cm^2}$
and $\kappa_{c2}=0.1360 \un{mS/cm^2}$, there is an intermediate
regime $\kappa_{c1} < \kappa < \kappa_{c2}$ where discrete
transmission patterns $k:l$ occur. Here, $k$ is the number of spikes
generated at the initial node, while $l$ correspond to the number of
spikes transmitted to the final node. In principle, the coupling
parameter $\kappa$ could be tuned in such a way that any rational
transmission pattern $k:l$ is attained. This manifests as
generalized, delayed synchronization.

Note that a similar effect shows up when one drives the Hodgkin
Huxley system (i.e. a single node) with an ac-sinusoidal signal,
where the ratio of spiking events to driving periods exhibits a
rational number~\cite{aihara1984}.

\subsubsection*{Influence of channel noise}

\begin{figure}
\begin{center}
\epsfig{figure=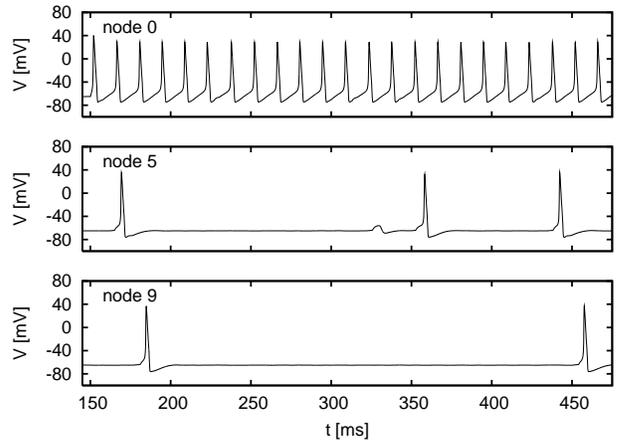, scale=0.67}
\caption[]{\label{fig:spiking} The simulated firing events vs. time, cf. Eqs.~\eqref{eq:comp_mod} -
\eqref{eq:gating}, \eqref{eq:stochasticgates}-\eqref{eq:noisecor}, 
of the noisy membrane potentials at three different nodes are depicted for 
$\kappa=0.065 \un{mS/cm^2}$ and the nodal area ${\cal A}=10^4 \un{\mu m^2}$. 
The constant current stimulus $I_{0, \mathrm{ex}}$ keeps the initial node ``0'' 
periodically firing. Noise-assisted spike propagation occurs only for some 
spike and some spikes do not reach the terminal node ``9''.  }   
\end{center}
\end{figure}

\begin{figure}[t!]
\begin{center}
\psfrag{A }{${\cal A}$}
\epsfig{figure=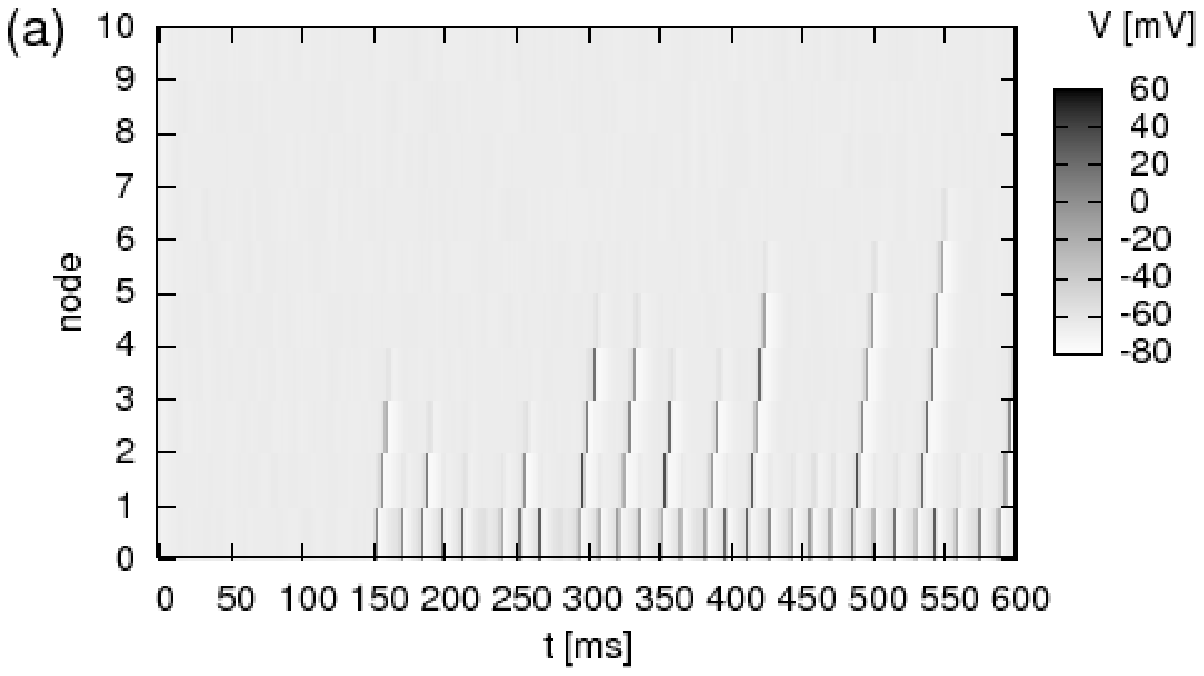,scale=0.67}\\
\epsfig{figure=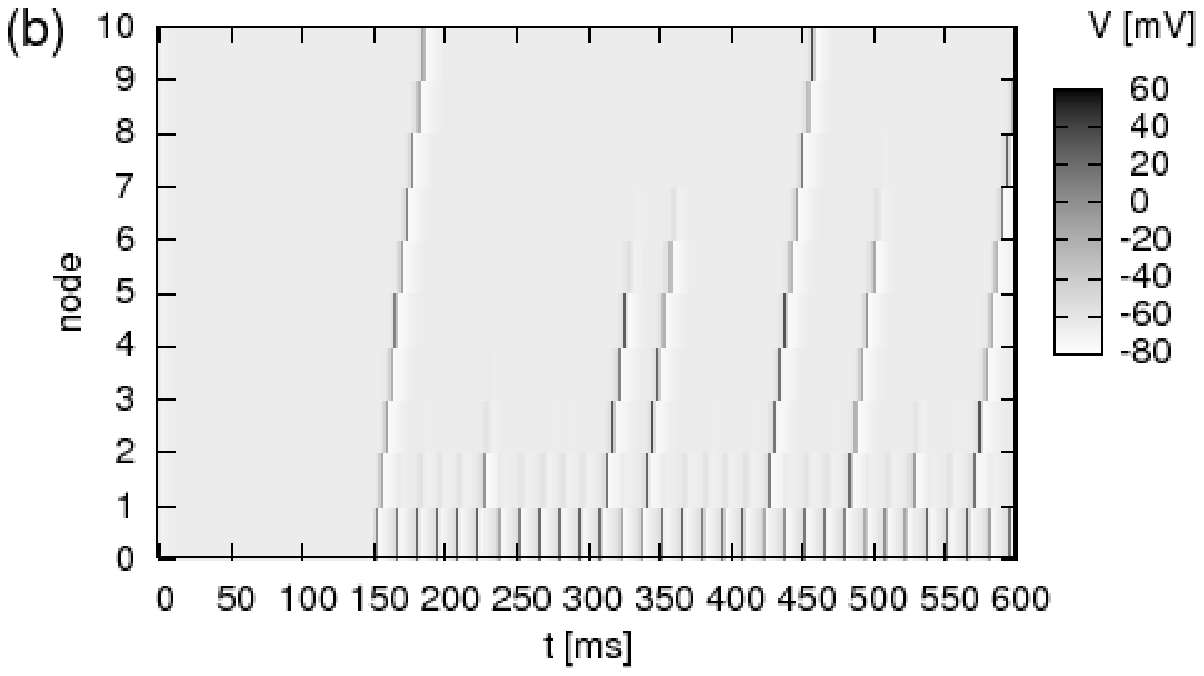,scale=0.67}\\
\epsfig{figure=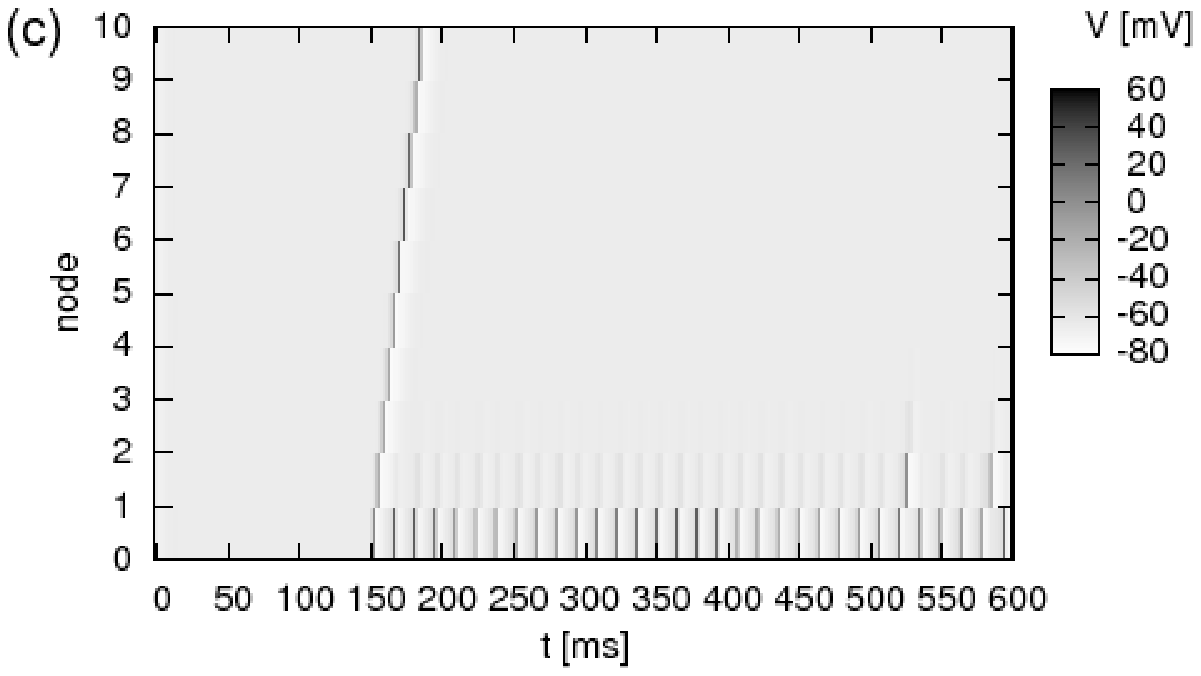,scale=0.67}\\
\caption[]{\label{fig:maps} Noisy spike propagation for a myelinated
neuron with $10$ Ranvier nodes and for $\kappa=0.065\ \mathrm{mS/cm^{2}}$. Single realizations  of
noise-assisted spike
  propagation for varying nodal membrane size ${\cal A}$. (a) ${\cal A}=250 \ \mathrm{\mu m ^2}$:  Strong channel noise
  inhibits strongly spike propagation   so that  propagation occurs only
  over a short distance. (b) ${\cal A}=10^4 \ \mathrm{\mu m ^2}$:
  An intermediate noise allows for the propagation over the longest
  distance. (c) ${\cal A}=5\times10^4 \ \mathrm{\mu m ^2}$: Weak channel
  noise only rarely allows to overcome the inter-nodal resistance.}
\end{center}
\end{figure}

\begin{figure*}[t]
\begin{center}
\psfrag{A }{${\cal A}$}
\epsfig{figure=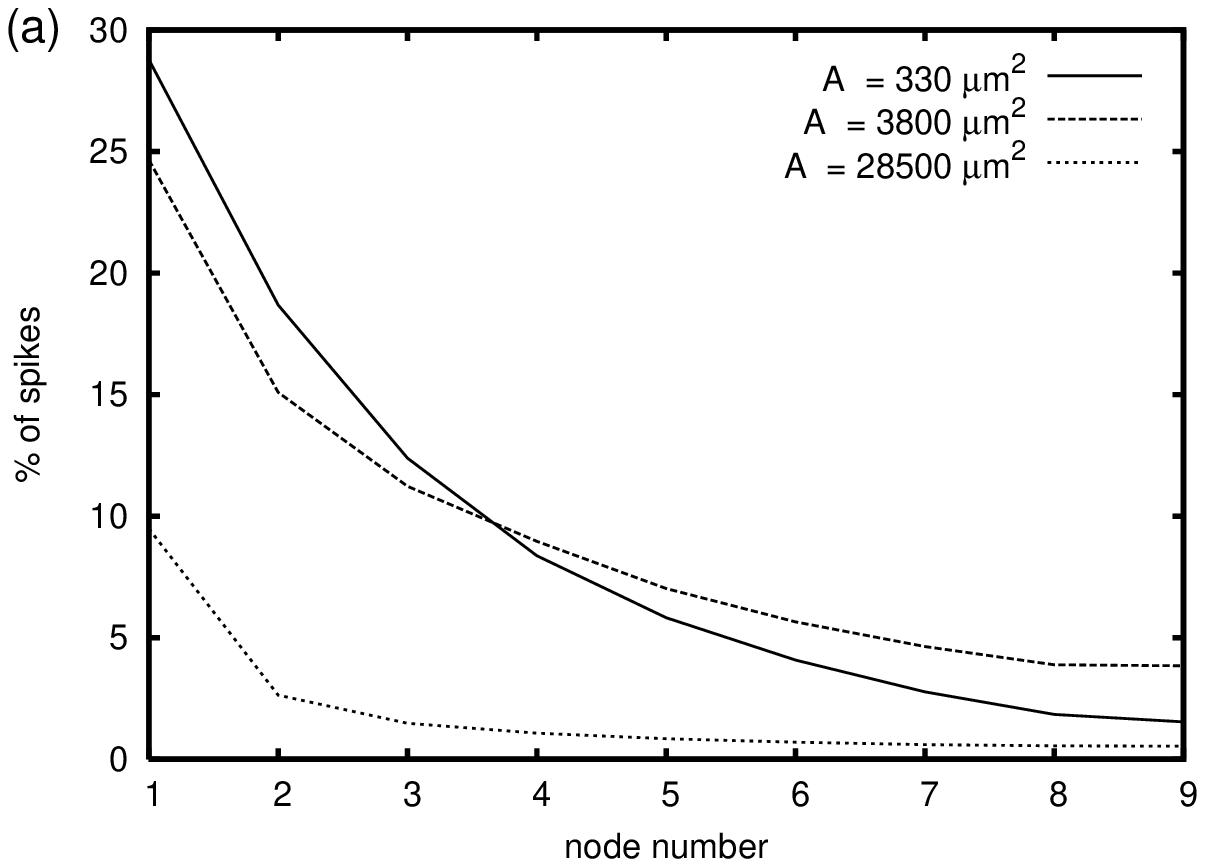,scale=0.6} 
\epsfig{figure=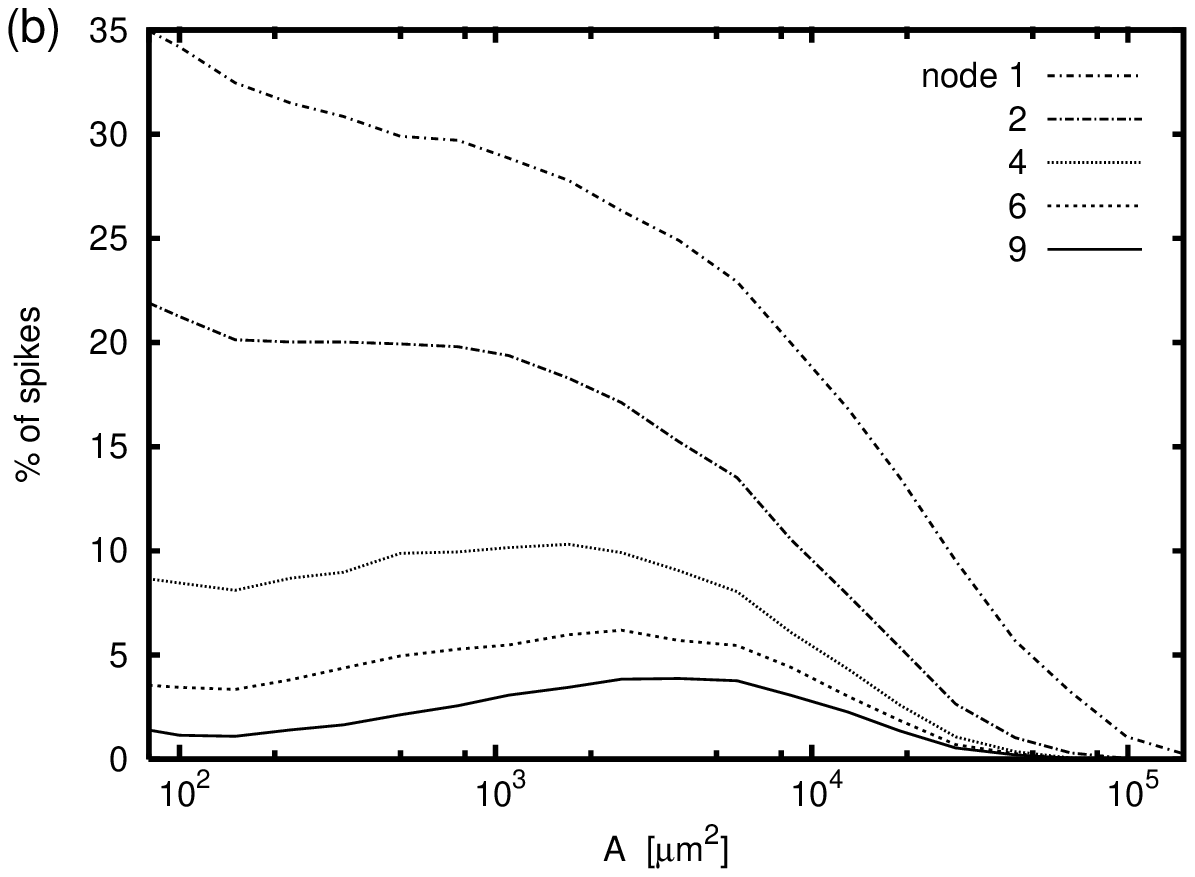,scale=0.6}\\
\caption[]{\label{fig:hist} Noise-assisted spike
  propagation for
  spikes initiated in the initial node (``0'') out of $10$ nodes of a linearly
  coupled chain: (a) The numerically obtained
  percentage of spike arrivals at the particular nodes and (b) its dependence on the nodal membrane
  size ${\cal A}$. The chosen
  coupling strength $\kappa=0.065\un{mS/cm^{2}}$ is sub-threshold. There occurs an
    optimum noise intensity (corresponding to a nodal membrane size ${\cal A}
    \approx 3800 \mathrm{\mu m^2}$) at which the signal transmission
    along the whole chain is most efficient. }
\end{center}
\end{figure*}

Next, we investigate how the nodal membrane size ${\cal A}$, determining
the channel noise intensity, influences the propagation of spikes
along the axon. Towards this objective we numerically integrated the
linearly coupled chain model with the nodes treated by the
stochastic Hodgkin-Huxley model, cf. Eqs.~\eqref{eq:comp_mod} -
\eqref{eq:gating} and
Eqs.~\eqref{eq:stochasticgates}-\eqref{eq:noisecor}.

Following the same protocol as for the deterministic case, we
computed the probability for a generated spike in the initial node to
be transduced to the terminal node. In our example the axon consists
again of $10$ nodes. For the coupling parameter, i.e. the
inter-nodal conductance, we chose a sub-threshold value of $\kappa =
0.065\un{mS/cm^{2}}$, i.e. below the critical value
$\kappa_{c1}$ for a deterministic spike propagation along the axon,
see Fig.~\ref{fig:kappa}. 
It turns out that for coupling parameters
slightly smaller than the critical value $\kappa_{c1}$ the influence
of the channel noise is most striking. Due to the channel noise,
 we observe spike propagation even for the {\itshape
  sub-threshold} coupling, i.e. for $\kappa < \kappa_{c1}$. The
intrinsic noise weakens the strict threshold and even for
sub-threshold values of the coupling parameter,
there is a non-vanishing probability for spike propagation.
In Fig.~\ref{fig:spiking} the spike train for three different nodes is 
shown for $\kappa = 0.065\un{mS/cm^{2}}$ and ${\cal A} = 10^{4}
\un{\mu m^{2}}$, i.e. for an intermediate intrinsic noise level. The
channel noise facilitates the propagation of action
potentials along the axonal chain. 

Fig.~\ref{fig:maps} depicts cases of  spike propagation in the
presence of noise of different intensities for the chosen
sub-threshold coupling parameter $\kappa$.
Apart from the noise facilitated spike propagation one can identify
another particular phenomenon: The failure of spike propagation
 due to a large  channel noise level. For rather strong intrinsic noise (for  small
nodal membrane sizes ${\cal A}$, yielding  small numbers of ion channels) the
spike initiated at the first node more likely propagates to the next
nodes. However, it is also likely that the spike collides with a
particularly large fluctuation and thus becomes deleted. Tracking
the behavior of a given node, one  observes the skipping of some
firing events~\cite{SchmidPhysicaA}, see Fig.~\ref{fig:spiking}. The propagation distance is
then quite short and the excitations rarely arrive at the terminal node,
see Fig.~\ref{fig:maps}(a). Note also, that for a high level of channel noise
spontaneous spikes can occur at any node. These {\itshape parasitic} spikes are not triggered by a
preceding spike in the neighboring nodes. Therefore, they
deteriorate  the information transfer along the axon and are not
considered for the spiking statistics at the terminal node.    

Too weak noise (i.e. a large number of channels) does not allow for
effective spike transmission to neighboring nodes. Only the very first excitation is
transmitted over a larger number of nodes, while the propagation of
the successive action potentials is rarely facilitated by noise, see
Fig.~\ref{fig:maps}(c). This specific enabling of propagation of the
first spike only is due to the initial setup of the problem (initial
conditions), because of the existence of a refractory time period.

For an intermediate channel noise intensity, however, the
propagation distance is the longest, although spikes propagate less
frequently than in the presence of an
intensive noise, see Fig.~\ref{fig:maps}(b). Overall, one may expect that there is an optimum dose of internal
noise for which the spike propagation along an axon is most
probable. To validate this assertion, we determined the probability
for an excitation stimulated in the initial node, to arrive at
the terminal node being which node ``9''in our case. 
The fraction of
  spikes arriving at a specific node depends on both, the distance
  which has to be covered and the noise
  level. With increasing distance, i.e. the node number, this fraction declines monotonically, cf. Fig.~\ref{fig:hist}(a). The
  decay depends on the channel noise level, i.e. the nodal membrane
  size ${\cal A}$. There is an
  intermediate noise level for which the decline is slowest.   
As a result, there exists an
optimum dose of internal noise (in our example for the nodal membrane
size near
${\cal A} \approx 3800 \mathrm{\mu m^2}$) at which the signal
transmission to the terminal node 
is most effective; i.e., the ratio of transmitted spikes assumes a maximum,
cf. Fig.~\ref{fig:hist}(b). 
We emphasize that the occurrence of the optimal dose of intrinsic noise
is robust under a change of the total number of nodes (not shown). 
Note that the observed maximum corresponds to a nodal number of
sodium ion channels of $N_{\chem{Na}}=2.28\cdot 10^{5}$ which
corroborates with the physiological range of the number found for
sodium ion channels in a node of Ranvier (experimental studies
report $N_{\chem{Na}} \approx 10^5$)~\cite{Sigworth,Conti}.

\section{\label{sec:concl}Conclusions}

The optimization of the signal propagation along a myelinated axon of
a finite size
occurs as the result of the competition between the constructive and
destructive role of channel noise occurring in the nodes of Ranvier.
On the one hand, when the number of channels in the node is very
small, the strong fluctuations of the activation potential
facilitate a spike propagation among nodes which otherwise does not
occur in the absence of channel noise. On the other hand, 
at a high level of
channel noise spikes cannot travel over a long distance because it
is likely that a spike soon collides with another noise-induced spike
leading to their mutual annihilation, or the noise can also suppress
spike generation. If the number of channels in nodes is large,
the fluctuations are too weak for a spike to overcome the internodal
resistance and subsequently to propagate. For a certain intermediate
number of channels in the node of Ranvier node, however, the intrinsic noise
is sufficiently strong  to induce the saltatory conduction while
being still sufficiently weak not to deteriorate the  spiking behavior.
An optimal nodal membrane size can be identified, for which the
signal becomes most efficiently transmitted over a certain axonal
distance. Moreover, the corresponding number of sodium ion channels
corresponds to  actual physiological values. This feature is quite in 
spirit of the stochastic resonance phenomenon
~\cite{gammaitoni,hanggi2002} with an intrinsic noise source
~\cite{Schmid,Jung}. One may therefore speculate, whether Nature
adopted this optimization method to balance the signal transmission
efficiency and the metabolic cost.


\acknowledgments This work has been supported by the Volkswagen
Foundation (project I/80424), the collaborative research center of
the DFG via SFB-486, project A-10 and the German Excellence
Initiative via the \textit {Nanosystems Initiative Munich} (NIM).


\begin{thebibliography}{00}
\bibitem{note} It is worth noticing that in our model study we use
  parameters for the squid giant axon. In realistic nodes of Ranvier,
  the sodium ion channel density is larger at least by factor 10 than
  that of the potassium ion channels. A physiologically more realistic
  model is left for a separate study.
\end{thebibliography}

\end{document}